\documentclass[superscriptaddress,twocolumn,notitlepage,showkeys,showpacs,longbibliography]{revtex4-1}
\usepackage[utf8]{inputenc}
\usepackage{stmaryrd}
\usepackage{amsmath}
\usepackage{amssymb}
\usepackage{graphicx}
\usepackage{dcolumn}
\usepackage{bm}
\usepackage{multirow}
\usepackage[colorlinks,linkcolor=blue,citecolor=blue,urlcolor=blue]{hyperref}
\usepackage{color}
\usepackage{ulem}
\usepackage{comment}
\usepackage{booktabs}

\begin{document}
\title{Abundant lattice instability in kagome metal ScV$_6$Sn$_6$}
\author{Hengxin Tan}
\author{Binghai Yan}
\affiliation{Department of Condensed Matter Physics, Weizmann Institute of Science, Rehovot 7610001, Israel}

\begin{abstract}
Kagome materials are emerging platforms for studying charge and spin orders. In this work, we have revealed a rich lattice instability in a $\mathbb{Z}_2$ kagome metal ScV$_6$Sn$_6$ by first-principles calculations.
Beyond verifying the $\sqrt3 \times \sqrt3 \times 3$ charge density wave (CDW) order observed by the recent experiment, we further identified three more possible CDW structures, i.e., $\sqrt3 \times \sqrt3 \times 2$ CDW with $P6/mmm$ symmetry, $2 \times 2 \times 2$ CDW with $Immm$ symmetry, and $2 \times 2 \times 2$ CDW with $P6/mmm$ symmetry. 
The former two are more energetically favored than the $\sqrt3 \times \sqrt3 \times 3$ phase, while the third one is comparable in energy. 
These CDW distortions involve mainly out-of-plane motions of Sc and Sn atoms, while V atoms constituting the kagome net are almost unchanged.
We attribute the lattice instability to the smallness of Sc atomic radius. In contrast, such an instability disappears in its sister compounds $R$V$_6$Sn$_6$ ($R$ is Y, or a rare-earth element) because $R$ has a larger radius.
Our work indicates that ScV$_6$Sn$_6$ might exhibit varied CDW phases in different experimental conditions and provides insights to explore rich charge orders in kagome materials. 

\end{abstract}

\maketitle

Kagome materials consisting of corner-shared-triangular atomic nets are in the frontier of research interests due to their diverse properties 
\cite{Ramirez1994strongly,Isakov2006hard,Ko2009doped,Guo2009,balents2010spin,yan2011spin,han2012fractionalized,Kiesel2012sublattice,Kiesel2013unconventional,wang2013competing,Nakatsuji2015Large,Yang2017topological,Liu2018Giant,ye2018massive,yin2018giant,Morali2019fermi,Liu2019magnetic,kang2020dirac}.
Electronic instabilities in kagome lattices had been predicted to occur at different electron filling and interactions \cite{Isakov2006hard,Kiesel2012sublattice,wang2013competing}.
The charge density wave (CDW) order was discovered in kagome superconductors $A$V$_3$Sb$_5$ ($A=$K, Rb, Cs) recently \cite{Ortiz2019new}.
The discovery of $A$V$_3$Sb$_5$ provides a unique platform to investigate the interaction between CDW \cite{jiang2021unconventional,Tan2021charge,Liang2021three,Li2021observation,Ortiz2021fermi,Park2021electronic,Wu2022charge,liu2022observation,scammell2023chiral}, superconductivity \cite{Ortiz2020Z2,Yin2021superc,chen2021roton,zhao2021cascade}, and topological band structure \cite{yang2020giant, Ortiz2020Z2, Fu2021quantum,hu2022rich,kang2022charge}.
On the other hand, the CDW was recently discovered in an antiferromagnetic kagome metal FeGe \cite{teng2022discovery,yin2022discovery}.
The CDW in FeGe provides extra freedom to study the interplay between CDW and magnetism \cite{setty2022electron,shao2022charge,teng2022intertwined,miao2022charge,zhou2022magnetic,wang2022first}, though the fundamental CDW structure remains to be resolved.
The interplay between different charge and spin orders in kagome materials is an emerging field with enriching kagome materials with various instabilities.

Very recently, Arachchige $et$ $al$ \cite{Arachchige2022charge} found a CDW phase transition with a wave-vector (1/3, 1/3, 1/3) at 92 K in the non-magnetic kagome metal ScV$_6$Sn$_6$, a new member of the versatile kagome family of the hexagonal HfFe$_6$Ge$_6$-type compounds which are studied extensively in magnetism \cite{venturini1991magnetic,zhang2020Topological,Ghimire2020Competing,Riberolles2022low,Rosenberg2022Uniaxial,Lee2022anisotropic,Pokharel2022highly,Zhang2022Electronic}, topology \cite{yin2020quantum,li2021dirac,Peng2021Realizing,xu2022topological,li2022manipulation,Hu2022Tunable}, and transport \cite{Asaba2020anomalous,ma2021rare,Gao2021anomalous,Roychowdhury2022large,Zeng2022large}.
Initial single crystal x-ray and neutron diffraction identified the CDW transition as the first-order \cite{Arachchige2022charge}, substantiated by a recent optical spectroscopy study \cite{hu2022optical}.
Different from the star-of-David or inverse star-of-David structural modulation involving mainly the in-plane displacement of the kagome sub-lattice in $A$V$_3$Sb$_5$ \cite{Tan2021charge}, the experiment revealed that the structural transition of ScV$_6$Sn$_6$ involves mainly the out-of-plane displacement of Sc and Sn atoms.
Besides, a recent high-pressure transport study finds that the CDW order can be maintained under pressure up to 2.4 GPa \cite{zhang2022destabilization}.
Being the first member showing charge instability in the diverse kagome family of HfFe$_6$Ge$_6$-type compounds, ScV$_6$Sn$_6$ opens the possibility of investigating the interplay between CDW and other electronic orders in these kagome materials.
However, basic knowledge about its electronic and structural properties is absent, and the origin of CDW transition in ScV$_6$Sn$_6$ remains unclear.

In this work, we have studied the electronic properties and structural origin of the CDW phase transition in ScV$_6$Sn$_6$ by first-principles calculations.
Our calculations revealed abundant lattice instabilities in ScV$_6$Sn$_6$.
In addition to verifying the observed $\sqrt3 \times \sqrt3 \times 3$ CDW structure, we reveal two more energy-favored CDW structures with different lattice modulations.
Unlike $A$V$_3$Sb$_5$, all CDW structures in ScV$_6$Sn$_6$ cannot quench the van Hove singularities near the Fermi energy, resulting in a limited decrease in the density of states (DOS).
Bare charge susceptibility calculations indicate that the Fermi surface nesting mechanism can be negligible to drive the CDW transition, consistent with Ref.~\onlinecite{hu2022optical}.
Additionally, CDW structures distinguished according to the crystallographic characteristics and band structures may inspire understanding of the CDW in FeGe, bearing their structural similarity. Our work predicted rich CDW phases in ScV$_6$Sn$_6$ that call for future experimental exploration.


\begin{figure}[tbp]
\includegraphics[width=0.95\linewidth]{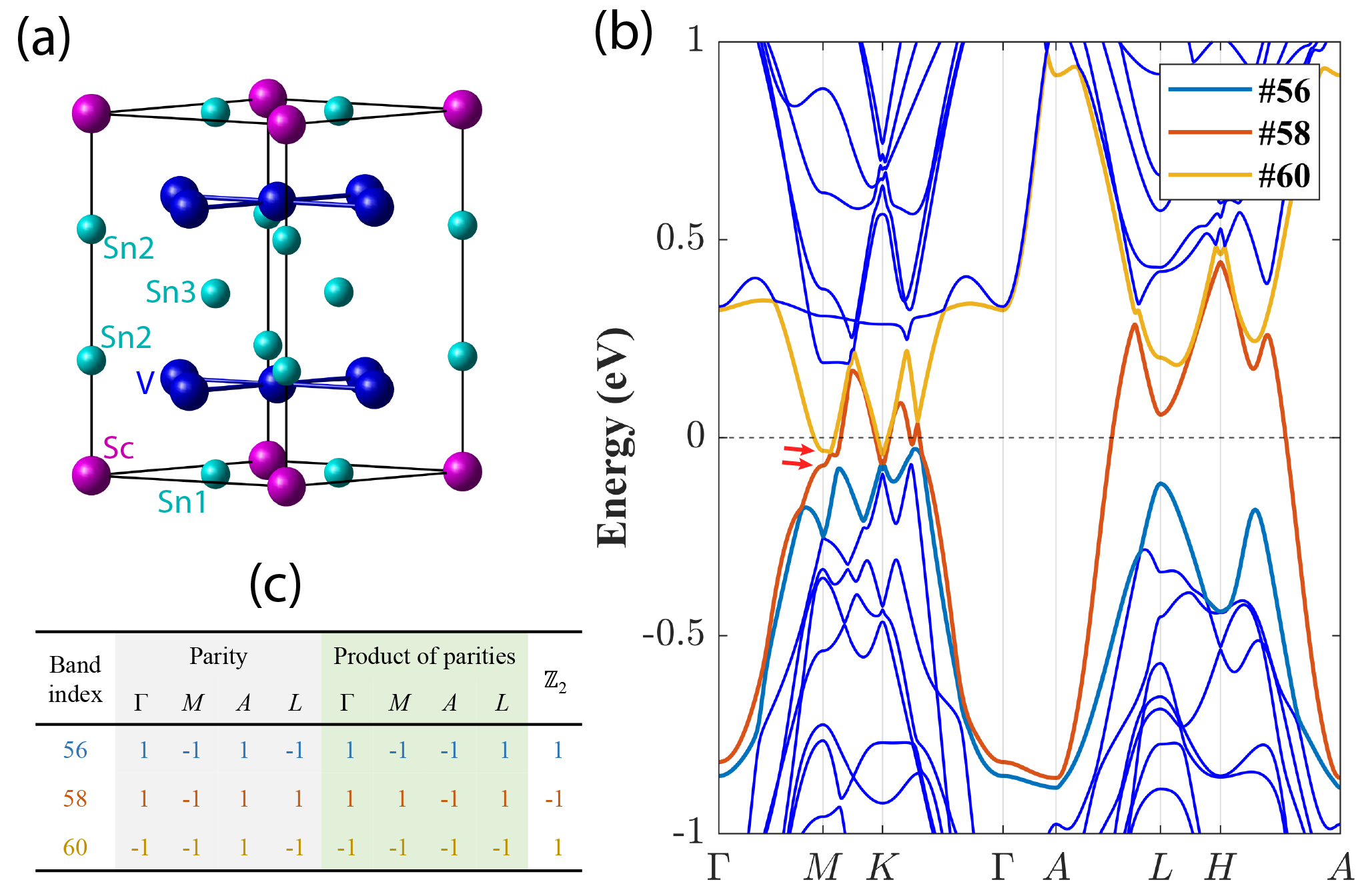}
\caption{\label{Fig1} (a) Crystal structure of the pristine ScV$_6$Sn$_6$. Non-equivalent atomic sites are labeled Sc, V, Sn1, Sn2, and Sn3. The V and Sn2 constitute the kagome layers; Sn3, the honeycomb layer; Sc and Sn1, the triangular layer.
(b) shows the electronic structure with spin-orbital coupling. Fermi energy is set to zero. Red arrows label the two van Hove singularities at the $M$ point.
(c) The parity of a given band [indicated in (b)], parity product up to this band, and the $\mathbb{Z}_2$-type topological invariant at the time-reversal invariant momenta.}
\end{figure}

The pristine structure of ScV$_6$Sn$_6$ is shown in Fig. \ref{Fig1}(a), which has the symmetry of space group $P6/mmm$ (Structural details are shown in Table S1 in Supplemental Material \cite{SM}.)
There is one non-equivalent V (Wyckoff position 6$i$), one non-equivalent Sc (1$a$), and three non-equivalent Sn atoms, Sn1 (2$c$), Sn2 (2$e$), and Sn3 (2d).
The structure comprises two kagome layers, one honeycomb layer, and a triangular layer.
The kagome layer comprises V and Sn2 atoms; the honeycomb layer comprises Sn3 atoms; and the triangular layer comprises Sc and Sn1 atoms.
Within each kagome layer, Sn2 atoms are slightly off the kagome sublayer formed by V.
Figure \ref{Fig1}(b) shows the band structure of the pristine phase.
Like typical kagome materials, there are van Hove singularities at the $M$ point and gapped Dirac points by spin-orbital coupling at the $K$ point near the Fermi energy.
A noticeable difference between ScV$_6$Sn$_6$ and $A$V$_3$Sb$_5$ is that Sn atoms in ScV$_6$Sn$_6$ do not contribute to the Fermi surface on the $k_z =0$ plane (see Fig. S8 in Supplemental Material \cite{SM}), while the Sb $p_z$ orbital contributes circular Fermi surfaces around $\Gamma$ and $A$ points in $A$V$_3$Sb$_5$.
The $\mathbb{Z}_2$ topological invariant \cite{Fu2007topological} is presented in Fig. \ref{Fig1} (c), which demonstrates the nontrivial topological nature of ScV$_6$Sn$_6$.

\begin{figure*}[tbp]
\includegraphics[width=0.97\linewidth]{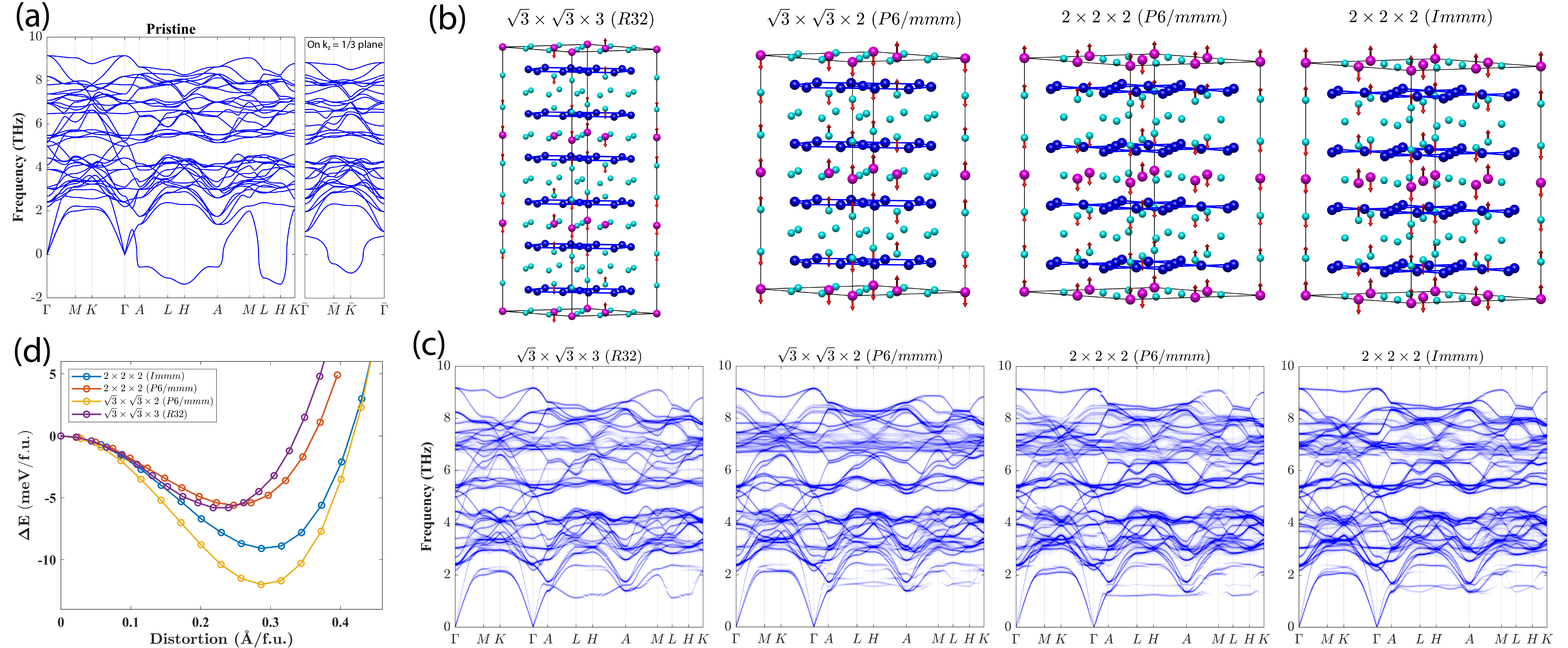}
\caption{\label{Fig2-phonon} (a) Phonon spectrum of the pristine ScV$_6$Sn$_6$. Left: conventional $k$ paths. Right: $k$ paths on the $k_z=1/3$ plane (in unit of $2\pi/c$) where $\bar \Gamma (0,0,1/3)$, $\bar M (1/2,0,1/3)$, $\bar K (1/3,1/3,1/3)$.
(b) Distortion patterns (red vectors) of four stable CDW structures with respect to the pristine phase in Fig \ref{Fig1}(a). The distortion mainly involves Sc and Sn2 atoms, while Sn1, Sn3, and V remain nearly unchanged. 
The conventional cell is used for the $\sqrt 3 \times \sqrt 3 \times 3$ ($R32$), and a hexagonal supercell is used for the $2 \times 2 \times 2$ ($Immm$) structures.
(c) Unfolded phonon spectrum of the four CDW structures in (b), which shows that the imaginary phonon branch on the $A-L-H$ plane in (a) is moved up to a favorable frequency regime, indicating the stability of these structures.
(d) Total energy of the CDW structures as a function of structural distortion amplitude [both are reduced to per formula unit (f.u.)].
}
\end{figure*}

To demonstrate the lattice instability, we performed phonon calculations based on the fully relaxed lattice structure. The phonon dispersion of pristine ScV$_6$Sn$_6$ is shown in Fig. \ref{Fig2-phonon}(a).
The entire $k_z = 1/2$ plane (in unit of 2$\pi$/$c$, $c$ is the out-of-plane lattice constant), i.e., the $A-L-H$ plane, shows imaginary phonon modes, and the lowest imaginary phonon is located at the $H$ point.
This behavior differs from the $A$V$_3$Sb$_5$ where only the $M-L$ line shows imaginary phonons in the pristine phase \cite{Tan2021charge}.
The multiple phonon instabilities on the $A-L-H$ plane indicate diverse structural instabilities.
Since the experimentally observed CDW has a wave-vector (1/3, 1/3, 1/3), we plot the phonon dispersion on the $k_z=1/3$ plane in Fig. \ref{Fig2-phonon}(a) as well.
This plane shows phonon instabilities on the $\bar M$(1/2, 0, 1/3)$-\bar K$(1/3, 1/3, 1/3) line, with the lowest imaginary phonon at the $\bar K$ point.
We emphasize that the global lowest imaginary phonon is at $H$ point.
To gain more insight into the structural distortion, we will focus on imaginary phonons at four $k$ points, i.e., $H$, $L$, $A$, and $\bar K$.

(i) $\bar K$(1/3, 1/3, 1/3). The imaginary phonon at $\bar K$ leads to a structural modulation of $\sqrt 3\times \sqrt 3 \times 3$.
We identified one stable structure after the complete structural relaxation by employing the imaginary phonon wave function at $\bar K$.
The $\sqrt 3\times \sqrt 3 \times 3$ structure has the $D_{3}$ symmetry (space group $R32$, No.155) as shown in Fig. \ref{Fig2-phonon}(b), which is the same as the experimentally reported CDW structure \cite{Arachchige2022charge}.
This structure can be further refined with $D_{3h}$ symmetry (space group $R \bar 3 m$, No.166), where the largest atomic displacement from the $R32$ structure is in the order of 10$^{-3}$ \AA.
Notice that the original experimental CDW structure can be readily refined with the $R \bar 3 m$ symmetry, where the most significant atomic movements are 0.002 \AA~ between $R \bar{3} m$ and $R32$.
Due to the negligible differences between the two structures, the electronic and phonon properties of the chiral $R32$ structure are almost identical to the central symmetric $R \bar 3 m$ structure, according to our calculations.
The phonon dispersion and total energy as a function of structural distortion of the $R32$ structure shown in Fig. \ref{Fig2-phonon}(c)\&(d) demonstrate the stability of two structures.

(ii) $H$(1/3, 1/3, 1/2). The imaginary phonon at $H$ point leads to a structural modulation of $\sqrt 3\times \sqrt 3 \times 2$ with the symmetry of $D_{6h}$ (space group $P6/mmm$, No. 191) as shown in Fig. \ref{Fig2-phonon}(b).
Its phonon dispersion (absence of imaginary mode) and total energy evolution in Figs. \ref{Fig2-phonon}(c) and \ref{Fig2-phonon}(d), respectively, indicate its stability.
Intriguingly, this structure has the lowest total energy among all CDW structures we identified, compatible with the lowest imaginary phonon at $H$.
We speculate this $\sqrt 3\times \sqrt 3 \times 2$ structure with $P6/mmm$ symmetry might exist as a stable phase under varied experimental conditions, which calls for future experimental verification.

(iii) $L$(1/2, 0, 1/2). The imaginary phonon at $L$ point leads to a structural modulation of $2\times2\times2$. Similarly, we identified two stable structures, one with $D_{6h}$ ($P6/mmm$, No.191) symmetry and the other with $D_{2h}$ symmetry ($Immm$, No.71), as shown in Fig. \ref{Fig2-phonon}(b). Their phonon dispersion and total energy evolution are shown in Fig. \ref{Fig2-phonon}(c)\&(d).

(iv) $A$(0, 0, 1/2). The imaginary phonon at $A$ point leads to a structural modulation of $1\times1\times2$. After full structural relaxation, we obtained one structure with $D_{6h}$ symmetry (space group $P6/mmm$). However, this structure still shows imaginary phonon modes at its Brillouin zone boundary, indicating additional instability in the basal plane, as shown by its phonon dispersion in Fig. S6 \cite{SM}.

Overall, we have identified four stable CDW structures in ScV$_6$Sn$_6$, including the experimentally reported one.
Their structural details are listed in Table S1 \cite{SM}.
The most energy-favored CDW structure has the $\sqrt 3\times \sqrt 3 \times 2$ lattice modulation. 
All these CDW structures involve mainly the out-of-plane movement of Sc and Sn2 atoms, while the movements of other atoms are negligible.
As shown in Fig. \ref{Fig2-phonon}(b), these different CDW structures can be obtained by different arrangements of the Sc and Sn2 atomic displacements.
We speculate more CDW structures might be established via such displacement arrangements, as evidenced by the multiple phonon instabilities in the $A-L-H$ plane. Multiple CDW phases may appear at varied conditions or coexist simultaneously in experiments. 

An emergent question is about the origin of CDW phase transition in ScV$_6$Sn$_6$.
In kagome lattices with van Hove singularity close to the Fermi energy, Fermi surface nesting is a potential mechanism for different charge orders.
As shown in Fig. \ref{Fig1}(b), two van Hove singularities are at $M$ near the Fermi energy.
The Fermi surface nesting function is estimated by the imaginary part of the bare charge susceptibility under zero-frequency limit,
\begin{equation}
\label{FSnesting}
\lim_{\omega \rightarrow 0} \chi''_0(\mathbf{q},\omega)/\omega = \sum_{nn',\mathbf{k}} \delta(\varepsilon_{n,\mathbf{k}} - \varepsilon_0) \delta(\varepsilon_{n',\mathbf{k}+\mathbf{q}} - \varepsilon_0),
\end{equation}
where $\delta$ is the Dirac-delta function. $\varepsilon_{n,\mathbf{k}}$ is the eigenvalue of band $n$ at $\mathbf{k}$ and $\varepsilon_0$ is the Fermi energy. $\mathbf{q}$ is the nesting vector. Generally, the Fermi surface nesting is responsible for the CDW phase transition only when both the imaginary and real parts of the bare charge susceptibility diverge at the same wave vector $\mathbf{q}$ (see discussions in, e.g., Ref. \onlinecite{Johannes2008fermi}). The real part defining the stability of charge density reads
\begin{equation}
\label{CDWfunc}
\chi'_0(\mathbf{q}) =  \sum_{nn',\mathbf{k}} \frac{\it{f}(\varepsilon_{n,\mathbf{k}}) - \it{f}(\varepsilon_{n',\mathbf{k}+\mathbf{q}})}{\varepsilon_{n',\mathbf{k}+\mathbf{q}} - \varepsilon_{n,\mathbf{k}}},
\end{equation}
where $f(\varepsilon)$ is the Fermi-Dirac distribution.

\begin{figure}[tbp]
\includegraphics[width=0.95\linewidth]{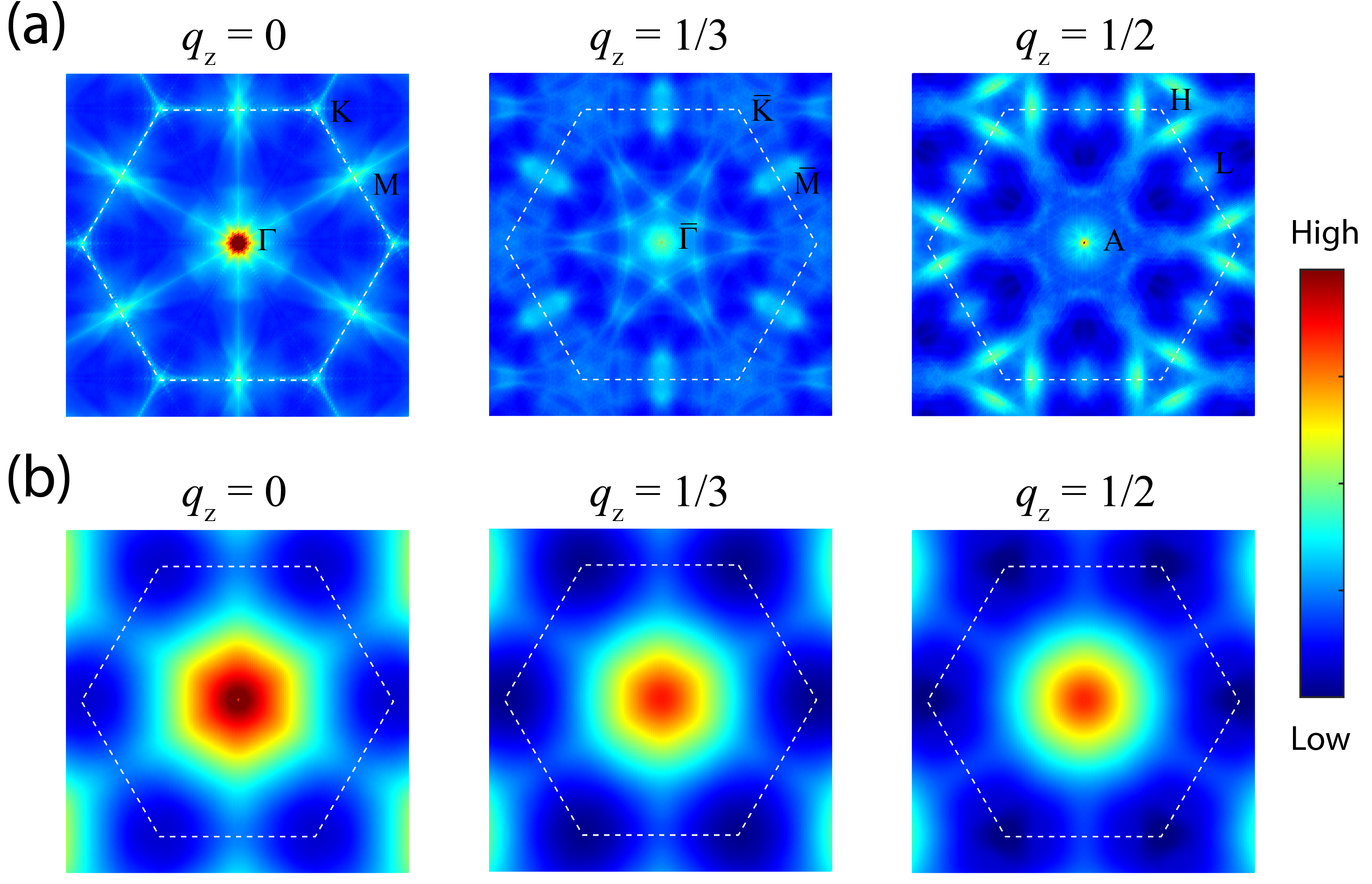}
\caption{\label{Fig3} Charge susceptibility of the pristine ScV$_6$Sn$_6$.
(a) shows the Fermi surface nesting function $\lim_{\omega \rightarrow 0} \chi''_0(\mathbf{q},\omega)/\omega$ as a function of $\mathbf{q}$ in the $q_z=0, \frac{1}{3}, \frac{1}{2}$ planes respectively. (b) is similar to (a) but for the real part $\chi'_0(\mathbf{q})$.
White hexagons in all panels represent the in-plane Brillouin zone.
}
\end{figure}

The imaginary and real parts of the bare charge susceptibility of the pristine ScV$_6$Sn$_6$ are plotted as a function of $\mathbf{q}$ in Fig. \ref{Fig3} (see Fig. S9 \cite{SM} for Fermi surfaces and additional susceptibility).
On all three $q_z = 0, \frac{1}{3}, \frac{1}{2}$ planes, the Fermi surface nesting function in Fig. \ref{Fig3}(a) shows small peaks at $M$, $\bar{M}$ and $L$ points respectively. The small peak at $K$ on the $q_z=0$ plane smears out at $\bar{K}$ on the $q_z = \frac{1}{3}$ plane and splits into three broadened peaks centered at $H$ on the $q_z = \frac{1}{2}$ plane. The peak at $M$ is compatible with the van Hove singularity-related Fermis surface nesting scenario.
However, the real part of the bare charge susceptibility $\chi'_0(\mathbf{q})$ in Fig. \ref{Fig3}(b) shows no peak near all these points.
This differs from the $A$V$_3$Sb$_5$, where the imaginary and real parts show small peaks simultaneously near the CDW $\textbf{q}$ vector (1/2, 0, 0) \cite{Wu2022charge}.
In addition, as shown by the unfolded band structures (Fig. S10) \cite{SM}, two van Hove singularities are unaffected by the CDW formation in ScV$_6$Sn$_6$.
Thus, we conclude the Fermi surface nesting mechanism does not necessarily drive the CDW formation in ScV$_6$Sn$_6$. This is consistent with the fact that V atoms in the kagome sublattice contributing to almost all Fermi surfaces remain nearly unchanged in positions during the CDW transition [Fig. \ref{Fig2-phonon}(b)].

We also checked the stability of YV$_6$Sn$_6$ \cite{Romaka2011Peculiarities}, an analog of ScV$_6$Sn$_6$ and found that YV$_6$Sn$_6$ shows no imaginary phonon band although YV$_6$Sn$_6$ exhibits a very similar band structure (Fig. S6 \cite{SM}).
The difference between YV$_6$Sn$_6$ and ScV$_6$Sn$_6$ comes from the larger effective radius of Y than Sc (1.02 vs. 0.87 \AA) \cite{shannon1969effective}.
As a result, the in-plane lattice constant of YV$_6$Sn$_6$ is expanded by about 1\% compared to ScV$_6$Sn$_6$ while the out-of-plane lattice constant $c$ is almost unchanged.
The larger in-plane space for Y and Sn2 releases the strain on Y/Sn2 for their out-of-plane movements, leading to more stability in YV$_6$Sn$_6$.
The radius of the Sc-site atom is thus critical for stabilizing the CDW phase in this material series.
This may explain why the CDW phase transition was rarely observed in rare earth atom-based isostructures $R$V$_6$Sn$_6$ ($R$ stands for rare earth elements and their effective ionic radii are similar to Y) \cite{Peng2021Realizing,Lee2022anisotropic,Ishikawa2021GdV6Sn6,Pokharel2021electronic,li2021dirac,Pokharel2022highly,Rosenberg2022Uniaxial,Zhang2022Electronic}.

\begin{figure}[tbp]
\includegraphics[width=0.95\linewidth]{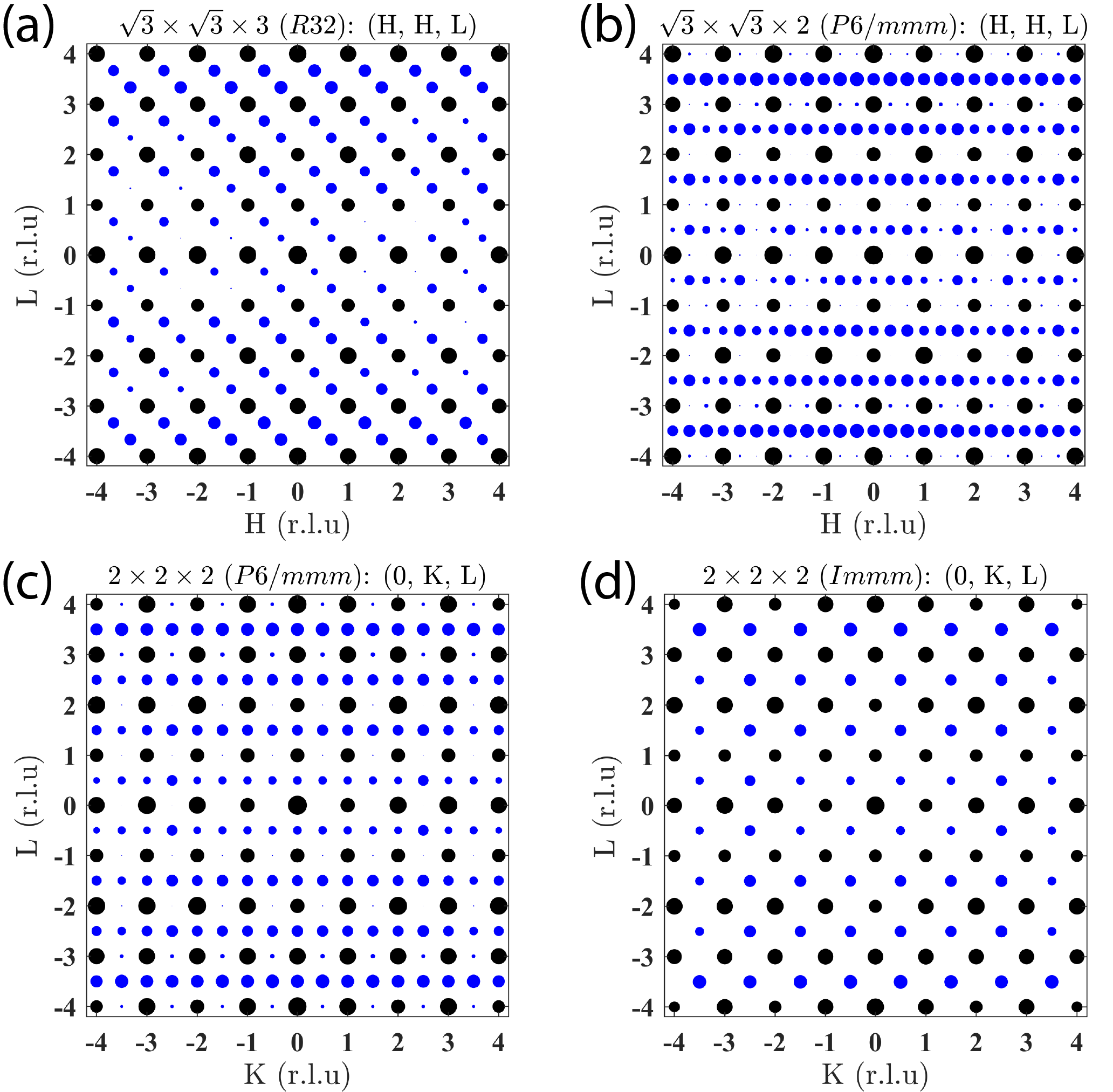}
\caption{\label{Fig4-XRD} XRD pattern of for (a) $\sqrt 3\times \sqrt 3 \times 3$ ($R32$), (b) $\sqrt 3\times \sqrt 3 \times 2$ ($P6/mmm$), (c) $2\times 2 \times 2$ ($P6/mmm$), and (d) $2\times 2 \times 2$ ($Immm$), respectively.
Black dots indicate Bragg peaks with integer Miller indices $H$, $K$, and $L$ (in reciprocal lattice unit of the pristine phase, r.l.u.).
Blue dots represent the CDW peaks absent in the pristine phase.
The dot size is proportional to the diffraction intensity.
}
\end{figure}

Another emergent question is distinguishing between these proposed CDW structures from the experiment.
Single crystal X-ray diffraction (XRD) is a powerful tool to distinguish between them from the crystal structure aspect. The XRD peak intensity is defined with $I(\bf{Q}) =  |F(\bf{Q})|^2$ where $F(\bf{Q})$ is the complex scattering amplitude,
\begin{equation}
    F(\textbf{Q}) = \sum_j f_j(\textbf{Q}) e^{ i \textbf{Q} \cdot \textbf{r}_j}.
    \label{FQ}
\end{equation}
\noindent The $\textbf{r}_j$ is the $j$th atomic position in the unit cell. $f_j(\textbf{Q})$ is the atomic form factor of the $j$th atom well approximated by a sum of Gaussians \cite{brown2006intensity}. The $\textbf{Q} = (H, K, L)$ stands for the momentum transfer in bases of the pristine reciprocal lattices ($H/K/L$ are not necessarily integers). The calculated XRD spectrum for the four CDW structures of ScV$_6$Sn$_6$ is shown in Fig. \ref{Fig4-XRD}.

As shown in Fig. \ref{Fig4-XRD}(a) and (b), the $\sqrt 3\times \sqrt 3 \times 3$ ($R32$) and $\sqrt 3\times \sqrt 3 \times 2$ ($P6/mmm$) CDW structures show CDW peaks at $H = \frac{1}{3}$ and $\frac{2}{3}$, as expected. Along the out-of-plane direction $L$, the $R32$ CDW structure shows also CDW peaks at $L = \frac{1}{3}$ and $\frac{2}{3}$, while the $P6/mmm$ CDW structure shows only CDW peaks at $L=\frac{1}{2}$. We notice that the experimental XRD pattern \cite{Arachchige2022charge} also shows CDW peaks at $(H,H,L) = (\frac{s}{3}+m, \frac{s}{3}+m, \frac{s}{3}+n)$ ($s=1,2$. $m$ and $n$ are arbitrary integers) while such CDW peaks are absent in Fig. \ref{Fig4-XRD}(a) due to a lattice reason. We speculate that there might be a 180$^\circ$ domain along the $c$ direction (i.e., the out-of-plane direction), which makes up the missing CDW peaks in Fig. \ref{Fig4-XRD}(a) and results in the observed XRD pattern (see section III.A in the Supplemental Material \cite{SM} for details).

For the two $2\times 2 \times 2$ CDW structures as shown in Fig. \ref{Fig4-XRD}(c) and (d), while they both display CDW peaks at the half-integer Miller indices, there are no CDW peaks at $(0,K,L)=(0,m,\frac{n}{2})$ and $(0,\frac{n}{2},m)$ of the $Immm$ phase ($m$ and $n$ are arbitrary integers). This difference originates from the differentiated lattice transformations from the pristine phase to the two CDW phases. A detailed derivation about the conditions for the presence/absence of the CDW peaks on different $(H,K,L)$ planes for all CDW structures is found in section III in the Supplemental Material \cite{SM}.

Electronic structures of the four CDW phases may also help distinguish the CDW structures.
The unfolded band structures (in the primitive Brillouin zone) of the four CDW structures are shown in Fig. S10 \cite{SM}.
The band on the $A-L-H$ plane mainly contributed by the Sn2 $p_z$ orbital and V $d_{xz}/d_{yz}$ orbitals from the kagome layer shows the largest changes by the CDW structural modulation.
Distinctly different from the $A$V$_3$Sb$_5$, the two van Hove singularities at $M$ near the Fermi energy are unaffected by all CDW formations in ScV$_6$Sn$_6$, which is in line with the recent optical spectroscopy study where no CDW gap is observed \cite{hu2022optical}.
As a result, the CDW formation in ScV$_6$Sn$_6$ does not induce much change to the DOS, in contrast to the apparent DOS decrease in $A$V$_3$Sb$_5$ upon CDW transition.
These characters can be rationalized by the fact that bands near the Fermi energy are mostly related to V, while the CDW distortion in ScV$_6$Sn$_6$ involves mainly the Sc and Sn2, which contribute little to the DOS.

The ScV$_6$Sn$_6$ (or the HfFe$_6$Ge$_6$-type kagome material) is closely related to the recently discovered magnetic kagome metal FeGe \cite{teng2022discovery,yin2022discovery} whose unknown CDW structure has been an obstacle in studying the electronic properties of the FeGe CDW phase.
The FeGe-type structure can be obtained by removing the Sc atom in ScV$_6$Sn$_6$ (the additional off-kagome-plane displacement of the Sn2 atoms is neglected).
Thus, in addition to making it feasible to study the interplay between charge instabilities and other exotic properties in the diverse HfFe$_6$Ge$_6$-type kagome family, the CDW in ScV$_6$Sn$_6$ may also assist in identifying the CDW structure of FeGe. For example, we identify a dynamically stable CDW structure of FeGe with $D_{2h}$ point group, a rotational symmetry-breaking phase, starting from the $2 \times 2 \times 2$ ($Immm$) CDW structure of ScV$_6$Sn$_6$ (Fig S11 \cite{SM}), which may help to understand the CDW and magnetic order in FeGe.

In summary, we have rationalized the experimentally observed CDW structure of the kagome metal ScV$_6$Sn$_6$ and identified three more possible CDW structures. 
The CDW phase transition in ScV$_6$Sn$_6$ is dominated by phonon softening, and the Sc-site atom radius is crucial for stabilizing the unique CDW structures.
Our work may indicate the delicate interaction between the charge instability and other exotic orders such as magnetism, topology, and correlation effect in HfFe$_6$Ge$_6$-type and FeGe-type kagome materials.
We hope our work will inspire more experiments to investigate the complex lattice instabilities of related kagome materials.

\textbf{Acknowledgement}. H.T. thanks Yongkang Li and Dr. Shangfei Wu for helpful discussions about single crystal X-ray diffraction and Dr. Yuji Ikeda for helpful discussions on phonon band unfolding. We also thank Prof. Stephen Wilson for fruitful discussions.
B.Y. acknowledges the financial support from the European Research Council (ERC Consolidator Grant ``NonlinearTopo'', No. 815869).

\textit{Note added.} During the review of our work, the predicted $\sqrt 3 \times \sqrt 3 \times 2$ CDW phase is observed as a short-ranged order by two recent inelastic X-ray scattering experiments \cite{cao2023competing,Korshunov2023softening}.


%

\end{document}


\title{Supplemental Material for ``Abundant lattice instability in kagome metal ScV$_6$Sn$_6$''}
\author{Hengxin Tan}
\author{Binghai Yan}
\affiliation{Department of Condensed Matter Physics, Weizmann Institute of Science, Rehovot 7610001, Israel}

\maketitle

\section{methodology}
\begin{figure*}[tbp]
\includegraphics[width=0.85\linewidth]{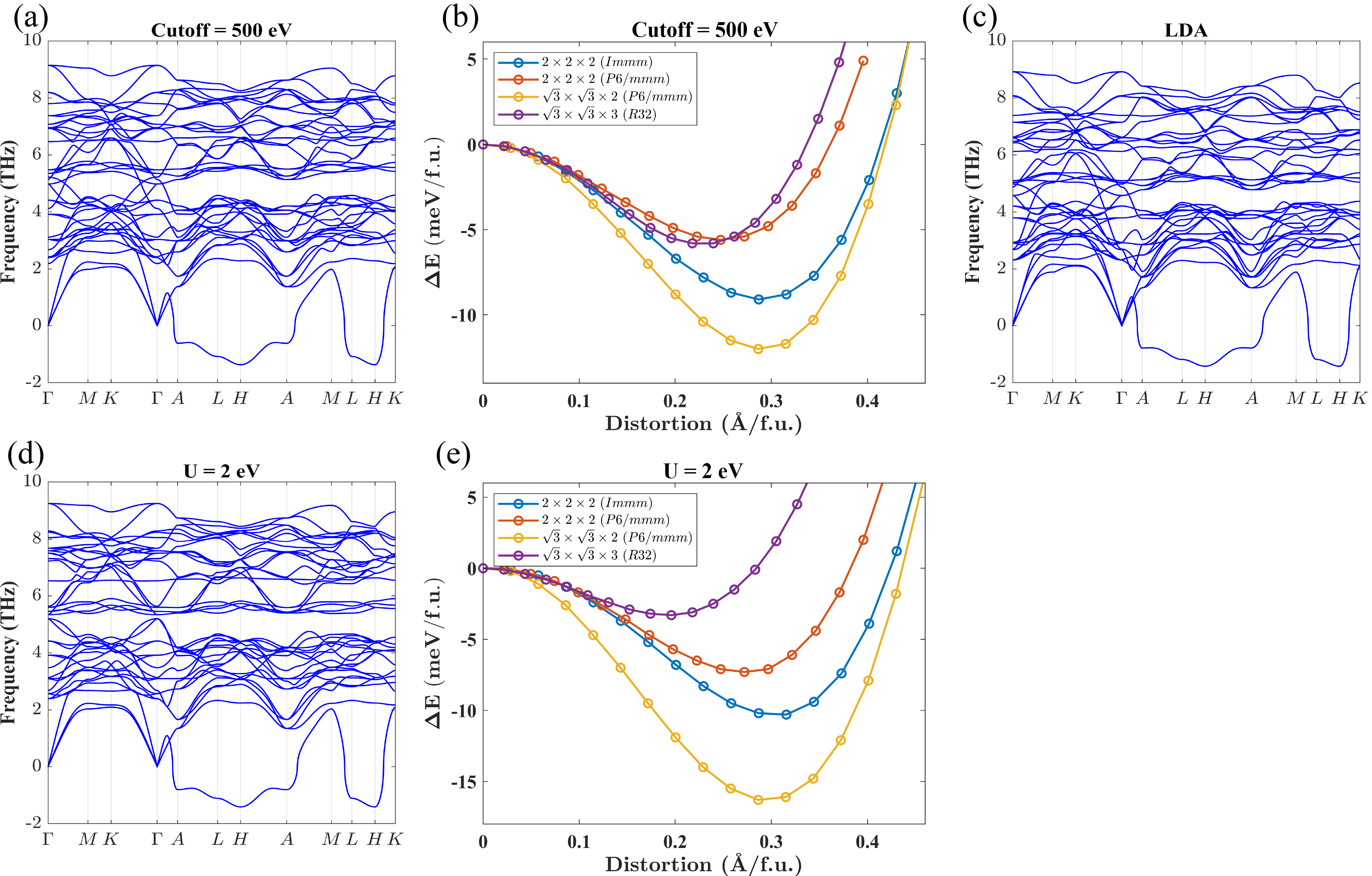}
\caption{\label{convergence_test} \textbf{Convergence tests of the calculations.}
(a) and (b) show the phonon dispersion of the pristine phase and total energy evolution of the CDW phases, respectively, as obtained from PBE calculations with an energy cutoff of 500 eV.
(c) displays the phonon band structure of the pristine phase under the local density approximation (LDA) for the electron-electron exchange-correlation interaction.
(d) and (e) are similar to (a) and (b) but under a Hubbard U (2 eV) correction for V $d$ and 300 eV energy cutoff.
}
\end{figure*}

All calculations are performed within the density functional theory (DFT) implemented in the plane-wave-based Vienna $ab$-$initio$ Simulation Package \cite{PRB54p11169,VASP}.
The generalized gradient approximation parameterized by Perdew, Burke, and Ernzerhof (PBE) \cite{PRL77p3865} is used to mimic the exchange-correlation interaction between electrons throughout.
The PBE-type projector augmented wave potentials with 3 valence electrons (3$d^1$4$s^2$) for Sc, 5 valence electrons (3$d^3$4$s^2$) for V, and 4 valence electrons (5$s^2$5$p^2$) for Sn are employed.
The energy cutoff for the plane wave basis set is 300 eV, which has been tested to converge all results [see Fig. \ref{convergence_test}(a)-(b)].
Atomic positions of all structures are fully relaxed (without spin-orbital coupling) until the remaining forces on atoms are less than 10$^{-3}$ eV/\AA, where $k$-meshes of reciprocal space resolution of 2$\pi\times$0.016 \AA$^{-1}$~are used for all structures.
Lattice constants of the pristine phase are also fully relaxed, while lattice constants of CDW structures are obtained from the pristine phase and kept fixed in structural relaxation.
In self-consistent calculations for all structures (with spin-orbital coupling), $k$-meshes are employed so that the reciprocal space resolution is no larger than 2$\pi\times$0.011 \AA$^{-1}$ (21$\times$21$\times$10 for pristine phase). The Gaussian smearing method of 0.1 eV is employed in all calculations.
The Fermi surface is calculated with the tight-binding Hamiltonian obtained from the Wannier90 code \cite{wannier90}, where Sc $d$, V $d$, and Sn $p$ orbitals are employed as the initial wannier basis set.
In the bare charge susceptibility calculation, a $k$-mesh of 120$\times$120$\times$60 is employed. The temperature in the Fermi-Dirac distribution function is about 116 K ($\sim$10 meV), and the delta function is represented by a Lorentzian function with a half-height width of 10 meV.

Phonon dispersions are calculated (without spin-orbital coupling) using the finite displacement method implemented in the \textsc{phonopy} software \cite{phonopy}, and unfolded with the \textsc{upho} software \cite{Yuji2017mode}.
\begin{itemize}
\item For the pristine phase, we employ a 3$\times$3$\times$3 supercell (351 atoms) combined with a $k$-mesh of 4$\times$4$\times$3. Our tests show that a supercell of 2$\times$2$\times$4 (208 atoms) with a 6$\times$6$\times$3 $k$-mesh can also expose structural instability, while a 2$\times$2$\times$1 supercell is insufficient.
\item For the $\sqrt 3 \times \sqrt 3 \times 3$ ($R32$) CDW structure, a supercell of 2$\times$2$\times$2 with a $k$-mesh of 3$\times$3$\times$3 is used.
\item For the $\sqrt 3 \times \sqrt 3 \times 2$ ($P6/mmm$) CDW structure, a supercell of 1$\times$1$\times$2 with a $k$-mesh of 7$\times$7$\times$3 is employed.
\item For the $2\times 2 \times 2$ ($P6/mmm$) CDW structure, a supercell of 1$\times$1$\times$2 with a $k$-mesh of 6$\times$6$\times$3 is employed.
\item For the $2 \times 2 \times 2$ ($Immm$) CDW structure, we first calculate the phonon dispersion of its hexagonal supercell (i.e., a $2\times 2 \times 2$ supercell of the pristine structure) with similar settings to the $2\times 2 \times 2$ ($P6/mmm$) CDW structure phonon calculation and then unfold the phonon dispersion back to the primitive cell of the $2 \times 2 \times 2$ ($Immm$) CDW structure.
\end{itemize}

We also have tested the effect of exchange-correlation interaction and a Hubbard U correction for V $d$ on the charge instabilities of ScV$_6$Sn$_6$, which also show robust charge instability of the pristine phase and robust relative stability of CDW structures. The results are shown below in Fig. \ref{convergence_test}(c)-(e).
Because all methods predict similar results and conclusions are the same, we employ the PBE results in both the main text and the following discussions.

\newpage
\section{Crystal structure of ScV$_6$Sn$_6$}

Structural details (supercell size, lattice parameters, space group, and Wyckoff positions) of the pristine and CDW phases of ScV$_6$Sn$_6$ are shown in Table \ref{Table_str}.
The $\sqrt 3 \times \sqrt 3 \times 3$ ($R\bar3m$) CDW is obtained by refining further the $\sqrt 3 \times \sqrt 3 \times 3$ ($R32$) CDW structure with the $R\bar3m$ symmetry, which shows the very small structural difference.
The lattice constants of all CDW structures are a rigid multiple of the pristine structure.
All these CDW structures are dynamically stable and energetically favorable.

\begin{table}
\centering
\caption{\label{Table_str} Lattice parameters ($a,b,c$ and $\alpha, \beta, \gamma$) and Wyckoff positions of the pristine and predicted CDW structures. Notice that the $\sqrt 3 \times \sqrt 3 \times 3$ ($R\bar3m$) CDW structure shows very small difference with the $\sqrt 3 \times \sqrt 3 \times 3$ ($R32$) CDW structure.
The conventional cells for CDW structures of $\sqrt 3 \times \sqrt 3 \times 3$ ($R\bar3m$), $\sqrt 3 \times \sqrt 3 \times 3$ ($R32$), and $2 \times 2 \times 2$ ($Immm$) are different from their respective primitive cells (minimum periodic cell) used in Sec. \ref{XRD}.
}
\renewcommand\arraystretch{1.1}
\begin{ruledtabular}
\begin{tabular}{l|l|l} \label{tbA1}
  Pristine phase ($P6/mmm$) & CDW $\sqrt 3 \times \sqrt 3 \times 3$ ($R32$) & CDW $\sqrt 3 \times \sqrt 3 \times 3$ ($R\bar3m$) \\
  \hline
  $a = b = 5.451 $\AA,  $c = 9.241$ \AA           & $a=b=9.441$ \AA, $c = 27.724$ \AA                & $a=b=9.441$ \AA, $c = 27.724$ \AA \\
  $\alpha = \beta = 90^\circ$, $\gamma=120^\circ$ & $\alpha = \beta = 90^\circ$, $\gamma=120^\circ$  & $\alpha = \beta = 90^\circ$, $\gamma=120^\circ$  \\
  &&\\
  Sc1 1a (0, 0, 0)         & Sc1 3a  (0, 0, 0)                   & Sc1 3a \ (0, 0, 0)                \\
  V1 \ 6i (0.5, 0, 0.24774)& Sc2 6c  (0,0, 0.32748)              & Sc2 6c \ (0, 0, 0.67252)          \\
  Sn1 2c (1/3, 2/3, 0)     & V1  18f (0.83368, 0.66741, 0.25065) & V1 18h (0.83296, 0.16704, 0.08268)\\
  Sn2 2e (0, 0, 0.31913)   & V2  18f (0.99978, 0.49987, 0.25087) & V2 18h (0.16656, 0.83344, 0.08246)\\
  Sn3 2d (1/3, 2/3, 0.5)   & V3  18f (0.16622, 0.83381, 0.25083) & V3 18h (0.50046, 0.49954, 0.08251)\\
                           & Sn1 6c  (0, 0, 0.89215)             & Sn1 6c \ (0, 0, 0.43382)          \\
                           & Sn2 6c  (0, 0, 0.22243)             & Sn2 6c \ (0, 0, 0.10785)          \\
                           & Sn3 6c  (0, 0, 0.56618)             & Sn3 6c \ (0, 0, 0.77757)          \\
                           & Sn4 9d  (0.66702, 0, 0)             & Sn4 18f (0.66702, 0, 0)           \\
                           & Sn5 9d  (0.33300, 0, 0)             & Sn5 18g (0.66783, 0, 0.5)         \\
                           & Sn6 9e  (0.66783, 0, 0.5)           &                                   \\
                           & Sn7 9e  (0.33213, 0, 0.5)           &                                   \\
\hline
\hline
\specialrule{0em}{5pt}{3pt}
\hline
\hline
 CDW $\sqrt 3 \times \sqrt 3 \times 2$ ($P6/mmm$) & CDW $2 \times 2 \times 2$ ($P6/mmm$) & CDW $2 \times 2 \times 2$ ($Immm$)\\
 \hline
  $a = b = 9.441 $\AA,  $c = 18.483$ \AA          & $a=b=10.902$ \AA, $c = 18.483$ \AA               & $a=5.451$ \AA, $b=9.441$ \AA, $c = 18.483$ \AA \\
  $\alpha = \beta = 90^\circ$, $\gamma=120^\circ$ & $\alpha = \beta = 90^\circ$, $\gamma=120^\circ$  & $\alpha = \beta = \gamma = 90^\circ$  \\
  &&\\
   Sc1 4h  (1/3, 2/3, 0.74054)        & Sc1 6i (0.5, 0, 0.75800)          & Sc1 4i (0, 0, 0.74032) \\
   Sc2 2e  (0, 0, 0.76014)            & Sc2 2e (0, 0, 0.74063)            & V1 16o (0.74771, 0.74971, 0.12615)\\
   V1 12o  (0.83370, 0.66741, 0.87406)& V1 12o (0.75050,0.50101,0.87396)  & V2 4j  (0.5, 0, 0.62575)\\
   V2 12o  (0.83294, 0.66588, 0.37369)& V2 12o (0.74954,0.49908,0.37371)  & V3 4j  (0.5, 0, 0.12660)\\
   V3  6i  (0.5, 0, 0.87373)          & V3 12n (0.74921,0,0.87356)        & Sn1 8l (0, 0.66663, 0.24948)\\
   V4  6i  (0.5, 0, 0.37404)          & V4 12n (0.75073,0,0.37417)        & Sn2 4i (0, 0, 0.58285)\\
   Sn1 12n (0.66660, 0, 0.74898)      & Sn1 12o (0.83333,0.66666,0.74988) & Sn3 4i (0, 0, 0.09965)\\
   Sn2  4h (1/3, 2/3, 0.90048)        & Sn2 4h (1/3, 2/3, 0.75088)        & Sn4 4g (0, 0.66880, 0)\\
   Sn3  4h (1/3, 2/3, 0.41684)        & Sn3 6i (0.5, 0, 0.91593)          & Sn5 4h (0, 0.33522, 0.5)\\
   Sn4  2e (0, 0, 0.91744)            & Sn4 6i (0.5, 0, 0.40219)          & \\
   Sn5  2e (0, 0, 0.39957)            & Sn5 2e (0, 0, 0.90045)            & \\
   Sn6  6j (0.66514, 0, 0)            & Sn6 2e (0, 0, 0.41694)            & \\
   Sn7  6k (0.66825, 0, 0.5)          & Sn7 6l (0.83427, 0.66853, 0)      & \\
                                      & Sn8 2c (1/3, 2/3, 0)              & \\
                                      & Sn9 6m (0.83239, 0.66478, 0.5)    & \\
                                      & Sn10 2d (1/3, 2/3, 0.5)           & \\
\end{tabular}
\end{ruledtabular}
\end{table}


\newpage
\section{Single crystal X-ray diffraction} \label{XRD}
In this section, we will define the lattice transformation matrix for each CDW phase and derive the single crystal X-ray diffraction reflection conditions.
Set $A_0$ and $A$ the lattice matrix of the pristine phase and the primitive cell of the CDW phase. Then the relationship between them is defined by the lattice transformation matrix $M$
\begin{equation}
    A = M \cdot A_0.
\end{equation}
\noindent Notice that the $M$ matrix depends on how the primitive structure of the CDW phase is taken (or equivalently, $A$ depends on how the smallest periodic cell of the CDW phase is chosen).

The single crystal X-ray diffraction intensity follows $I(\textbf{Q}) = \lvert F(\textbf{Q}) \rvert^2$. $\lvert F(\textbf{Q}) \rvert $ is the complex scattering amplitude
\begin{equation}
    F(\textbf{Q}) = \sum_j f_j(\textbf{Q}) e^{ i \textbf{Q} \cdot \textbf{r}_j}
    \label{FQ}
\end{equation}
\noindent where $\textbf{r}_j$ is the $j$th atomic position and the $\textbf{Q} = (h, k, l)$ is the momentum transfer in reciprocal space, based on the reciprocal lattice vectors of $A_0$ or $A$.
In the CDW experiment, $\textbf{Q}$ is generally based on the pristine phase $A_0$.
In the following, the Miller indices $\textbf{Q}_0 = (H, K, L)$ and $\textbf{Q} = (h, k, l)$ are used for the pristine and CDW phases, respectively.
The relationship between the $\textbf{Q}_0 = (H, K, L)$ and $\textbf{Q} = (h, k, l)$ is derived as
\begin{equation}
\begin{aligned}
    \textbf{Q}_0 &= \textbf{Q} \cdot (M^T)^{-1}
\end{aligned}
\end{equation}
Notice that the $(h, k, l)$ of the CDW phase are all integers, while $(H, K, L)$ don't have to be integers.
In the following, the diffraction peaks at $(H, K, L)$ composed of all integers are considered Bragg peaks, while others are regarded as CDW peaks (the Bragg peak intensity can be modified by the CDW transition).
If in a certain plane, $(H, K, L)$ has only integer components, then no CDW peak can be observed at the relevant plane.

We emphasize that, even though the $M$ matrix depends on how the primitive structure of the CDW phase is taken as mentioned above, the X-ray diffraction pattern doesn't depend on $M$, once $A_0$ is fixed. Thus in the following analyses for the CDW structures, we take a typical $M$ matrix to derive characters of the X-ray diffraction pattern for each CDW without loss of generality.

\newpage

\subsection{CDW $\sqrt 3 \times \sqrt 3 \times 3$ ($R32$)}
For the $\sqrt 3 \times \sqrt 3 \times 3$ ($R32$) CDW phase, 
$$ M = \begin{bmatrix}
    \frac{2}{3} & \frac{1}{3} & \frac{1}{3}\\
    -\frac{1}{3} & \frac{1}{3} & \frac{1}{3}\\
    -\frac{1}{3} & -\frac{2}{3} & \frac{1}{3}
\end{bmatrix}
\begin{bmatrix}
    1 & -1 & 0\\
    1 & 2 & 0 \\
    0 & 0 & 3\\
\end{bmatrix}
= \begin{bmatrix}
    1 & 0 & 1\\
    0 & 1 & 1 \\
   -1 & -1 & 1\\
\end{bmatrix}
$$
The detailed relationship between $(H, K, L)$ and $(h, k, l)$ is
\begin{equation}
\begin{aligned}
    (H, K, L) &= (h, k, l) \cdot (M^T)^{-1}\\
    &= \frac{1}{3} (2h-k-l, -h+2k-l, h+k+l)
\end{aligned}
\end{equation}

(i) $L = N$ where $N$ is an integer, $h+k+l=3N$; then $(H, K, L) = (h-N, k-N, N)$;

(ii) $K = N$ where $N$ is an integer, $l=2k-h-3N$; then $(H, K, L) = (h-k+N, N, k-N)$;

(iii) $H = N$ where $N$ is an integer, $l=2h-k-3N$; then $(H, K, L) = (N, -h+k+N, h-N)$;

(iv) $H = K$, i.e., $h=k$; then $(H, K, L) = \frac{1}{3} (h-l, h-l, 2h+l)$;

The $(H, K, L)$ has only integer elements in cases (i)-(iii). Thus the X-ray diffraction pattern on these planes shows only Bragg peaks and no CDW peak.
To detect the CDW peaks, experiments should avoid a $(H, K, L)$ plane with a fixed integer for $H$ or $K$ or $L$.

Case (iv) $H = K$ could be used to identify the CDW peaks. But an additional reflection condition is applied for this case. If we further set $H=K=\frac{s}{3}+N$ where $s=1,2$ and $N$ is arbitrary integer, then we have $(H, K, L) = (\frac{s}{3},\frac{s}{3},\frac{2s}{3})+(N,N,2N+l)$. This means that the diffraction peaks can only appear at $(\frac{1}{3},\frac{1}{3},\frac{2}{3})$ or $(\frac{2}{3},\frac{2}{3},\frac{1}{3})$ (here the integer part of each element is ignored).
Thus no diffraction peak can appear at $(H, K, L) = (\frac{s}{3},\frac{s}{3},\frac{s}{3}) + (N,N,2N+l)$. In the experimental pattern, both sets of CDW peaks are observed. We speculate that there might be a 180$^\circ$ domain in the $c$ direction. In the 180$^\circ$ domain, the CDW peaks can only appear at $(\frac{s}{3},\frac{s}{3},-\frac{2s}{3})$ which is equivalent to $(\frac{s}{3},\frac{s}{3},\frac{s}{3})$ (integer part ignored) with $s=1,2$. Thus the two domains make up for each other's missing CDW peaks, resulting in the observed X-ray diffraction pattern.

\begin{figure*}[tbp]
\includegraphics[width=0.95\linewidth]{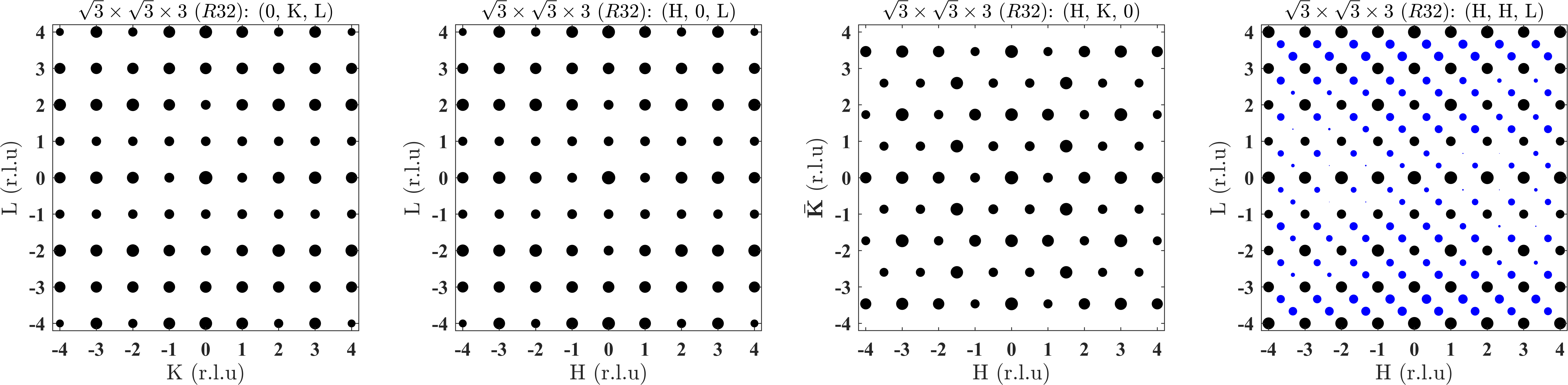}
\caption{\label{Fig-XRD-S1} \textbf{X-ray diffraction pattern of the $\sqrt 3\times \sqrt 3 \times 3$ ($R32$) CDW structure on different $k$-planes.}
The $H$, $K$, and $L$ indices are in reciprocal lattice units (r.l.u.) of the pristine phase.
Notice that for the $(H, K, 0)$ plane (the third panel), the $y$-axis does not correspond to the fractional $K$ due to the 60$^\circ$ angle between the two in-plane reciprocal lattice vectors.
The black dots with integer $H$, $K$, and $L$ are regarded as Bragg peaks while blue dots with at least one fractional $H$, $K$, or $L$ stand for CDW peaks.
Those black and blue dots are present and absent in the pristine phase, respectively.
The dot size is proportional to the diffraction intensity.
}
\end{figure*}

Relevant results can be found in Fig. \ref{Fig-XRD-S1}.

\newpage

\subsection{CDW $\sqrt 3 \times \sqrt 3 \times 2$ ($P6/mmm$)}
For the $\sqrt 3 \times \sqrt 3 \times 2$ ($P6/mmm$) CDW phase, 
$$ M = \begin{bmatrix}
    1 & -1 & 0\\
    1 & 2 & 0 \\
    0 & 0 & 2\\
\end{bmatrix}
$$
The relationship between $(H, K, L)$ and $(h, k, l)$ is
\begin{equation}
\begin{aligned}
    (H, K, L) &= (h, k, l) \cdot (M^T)^{-1}\\
    &= (\frac{2h+k}{3}, \frac{-h+k}{3}, \frac{l}{2} )
\end{aligned}
\end{equation}

Because the $L$ is independent of $H$ and $K$, whichever is an integer would not induce constraints to the others. Thus the CDW peak can appear on any $(H, K, L)$ plane. In our calculations for this CDW structure, the CDW peaks are very weak when $H$, $K$, and $L$ are close to zero, which is determined by the unique atomic structure involving only out-of-plane movements of atoms.

There are still some special cases that eliminate some CDW peaks. For example, for the case of $K=N$ ($N$ is an integer) plane, i.e., $k=h+3N$, we have $(H, K, L) = (h+N, N, \frac{l}{2})$. This indicates that the diffraction peak can only appear at integer $H$. As a result, such a $(H, K, L)$ plane can not be used to identify the in-plane $\sqrt 3 \times \sqrt 3$ lattice modulation. Similar discussions also apply to the case of $H=N$.

\begin{figure*}[tbp]
\includegraphics[width=0.95\linewidth]{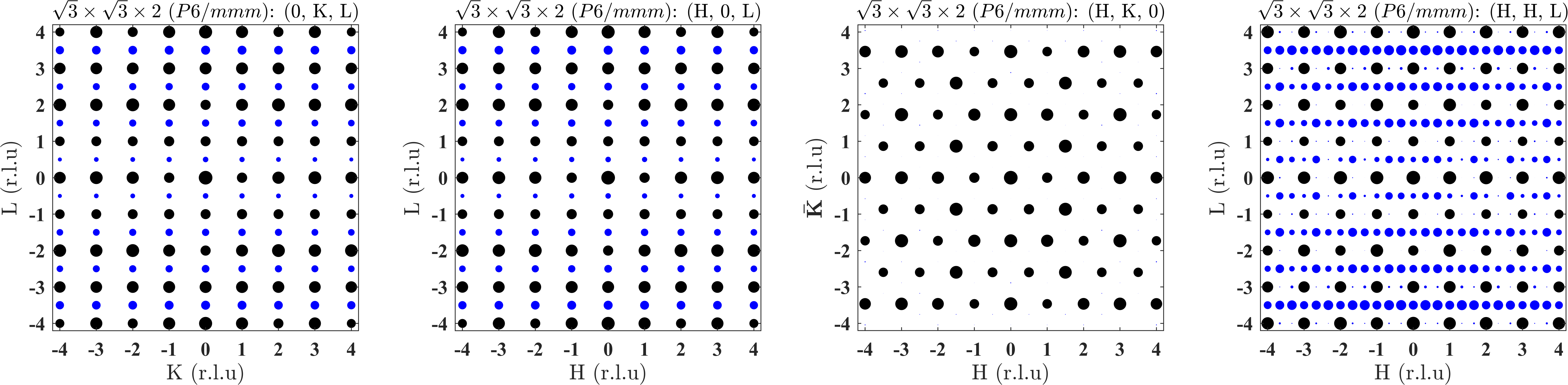}
\caption{\label{Fig-XRD-S2} Similar to Fig. \ref{Fig-XRD-S1} but for the CDW structure of $\sqrt 3 \times \sqrt 3 \times 2$ ($P6/mmm$).
}
\end{figure*}

Results are found in Fig. \ref{Fig-XRD-S2}.

\newpage

\subsection{CDW $2 \times 2 \times 2$ ($Immm$)}
For the $2 \times 2 \times 2$ ($Immm$) CDW phase, 
$$ M = \begin{bmatrix}
    0 & \frac{1}{2} & \frac{1}{2}\\
    0 & -\frac{1}{2} & \frac{1}{2}\\
    \frac{1}{2} & \frac{1}{2} & -\frac{1}{2}\\
\end{bmatrix}
\begin{bmatrix}
    2 & 0 & 0 \\
    0 & 2 & 0 \\
    0 & 0 & 2 \\
\end{bmatrix}
= \begin{bmatrix}
    0 & 1 & 1\\
    0 & -1 & 1\\
    1 & 1 & -1\\
\end{bmatrix}
$$
The relationship between $(H, K, L)$ and $(h, k, l)$ is
\begin{equation}
\begin{aligned}
    (H, K, L) &= (h, k, l) \cdot (M^T)^{-1}\\
    &= (k+l, \frac{h-k}{2}, \frac{h+k}{2})
\end{aligned}
\end{equation}

i) $L = N$ where $N$ is an integer, then $h+k=2N$; we have $(H, K, L) = (k+l, N-k, N)$;

ii) $K = N$ where $N$ is an integer, then $h-k=2N$, we have $(H, K, L) = (k+l, N, N+k)$;

iii) $H = K$, then $h=3k+2l$; we have $(H, K, L) = (k+l, k+l, 2k+l)$.

\begin{figure*}[tbp]
\includegraphics[width=0.95\linewidth]{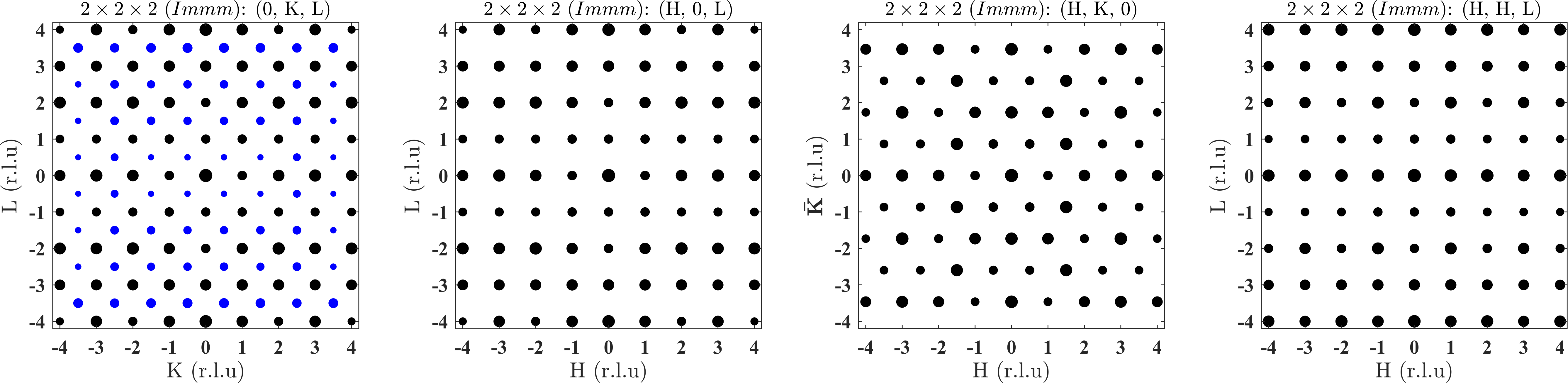}
\caption{\label{Fig-XRD-S3} Similar to Fig. \ref{Fig-XRD-S1} but for the CDW structure of $2 \times 2 \times 2$ ($Immm$).
}
\end{figure*}

For these cases, the $(H, K, L)$ has only integer elements inside. Thus in the X-ray diffraction pattern, only Bragg peaks are seen, and no additional CDW peak can be observed.
But for $H = N$ case where $(H, K, L) = (N, \frac{h-k}{2}, \frac{h+k}{2})$ can have half-integer elements, CDW peaks will appear. Thus this case could be used to show the CDW peaks. Be aware that, in this plane, no diffraction peaks can appear at $(H, K, L)=(N,N_1,\frac{N_2+1}{2})$ and $(N,\frac{N_1+1}{2},N_2)$ where $N_1$ and $N_2$ are arbitrary integers. Results are found in Fig. \ref{Fig-XRD-S3}.

As mentioned above, the lattice transformation matrix $M$ of the CDW phase can be defined differently. Thus the appearance of the reflection conditions may differ. But the physics behind it is robust and similar discussions can be performed for other choices of $M$.

\newpage

\subsection{CDW $2 \times 2 \times 2$ ($P6/mmm$)}
For the $2 \times 2 \times 2$ ($P6/mmm$) CDW phase, 
$$ M = \begin{bmatrix}
    2 & 0 & 0\\
    0 & 2 & 0 \\
    0 & 0 & 2\\
\end{bmatrix}
$$
The relationship between $(H, K, L)$ and $(h, k, l)$ is
\begin{equation}
\begin{aligned}
    (H, K, L) &= (h, k, l) \cdot (M^T)^{-1}\\
    &= (\frac{h}{2}, \frac{k}{2}, \frac{l}{2}).
\end{aligned}
\end{equation}
Whichever component of the $(H, K, L)$ is fixed to an integer, there can be half-integers for the other two components. Thus CDW peaks can appear on any plane. In our calculation for this CDW structure, the CDW peaks are very small when $H$, $K$, and $L$ are close to zero. This is related to the special atomic positions in this CDW structure, which can be obtained from Eq. \eqref{FQ}.

\begin{figure*}[tbp]
\includegraphics[width=0.95\linewidth]{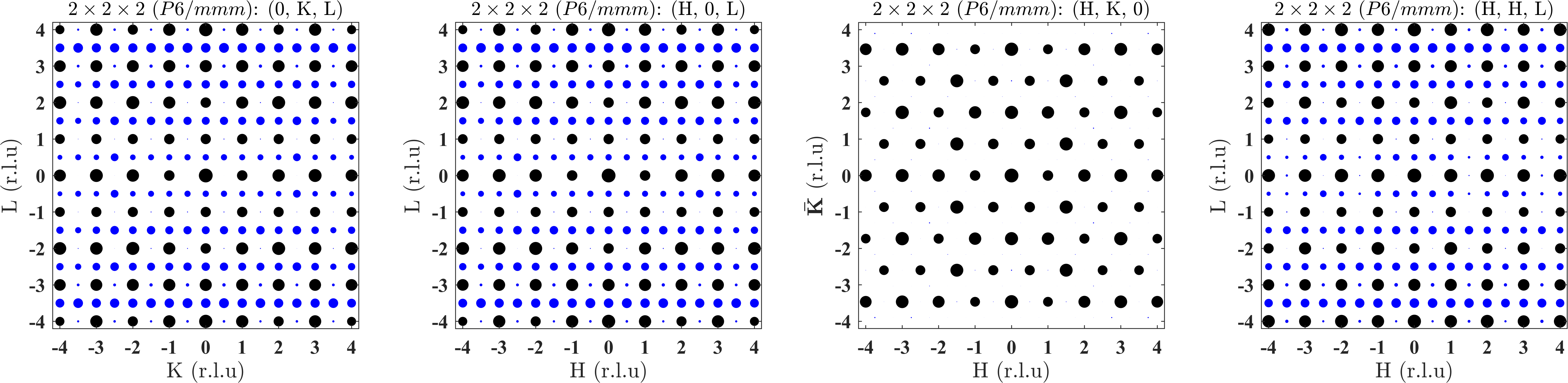}
\caption{\label{Fig-XRD-S4} Similar to Fig. \ref{Fig-XRD-S1} but for the CDW structure of $2 \times 2 \times 2$ ($P6/mmm$).
}
\end{figure*}

Results are found in Fig. \ref{Fig-XRD-S4}.

\newpage
\section{Other figures}
\begin{figure}[tbp]
\includegraphics[width=0.95\linewidth]{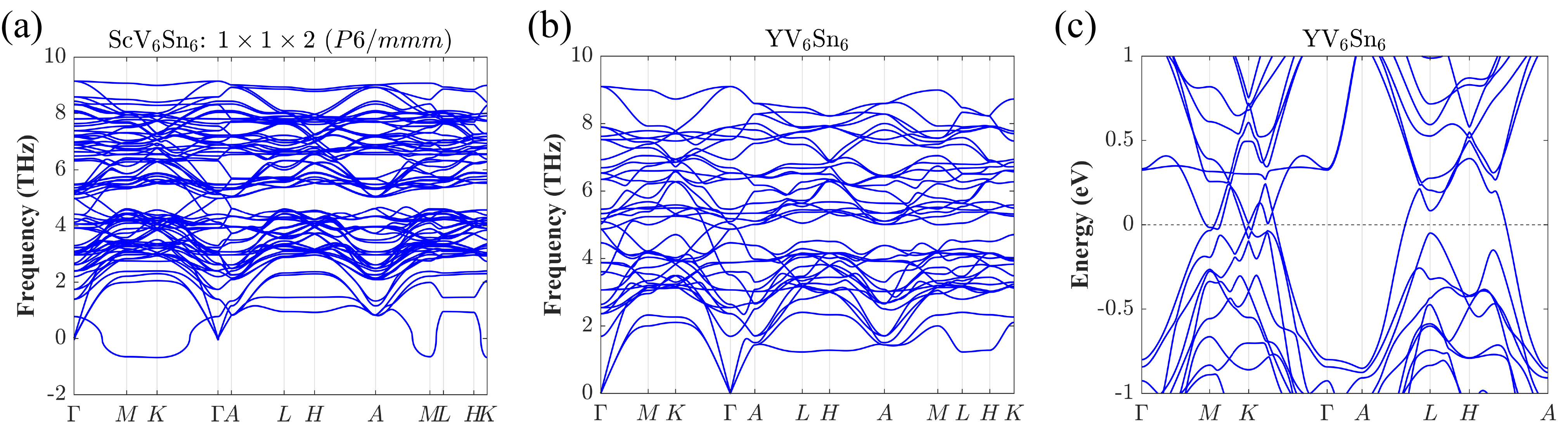}
\caption{ \label{FA1-112} Phonon band structure of (a) the $1 \times 1 \times 2$ structure ($P6/mmm$) obtained from the imaginary phonon at $A$(0, 0, 1/2) of ScV$_6$Sn$_6$, and (b) the pristine phase of YV$_6$Sn$_6$. (c) shows the electronic band structure of YV$_6$Sn$_6$ with spin-orbital coupling, which is very similar to that of the pristine ScV$_6$Sn$_6$ in Figure 1(b) in the main text.
}
\end{figure}

\begin{figure}[tbp]
\includegraphics[width=0.75\linewidth]{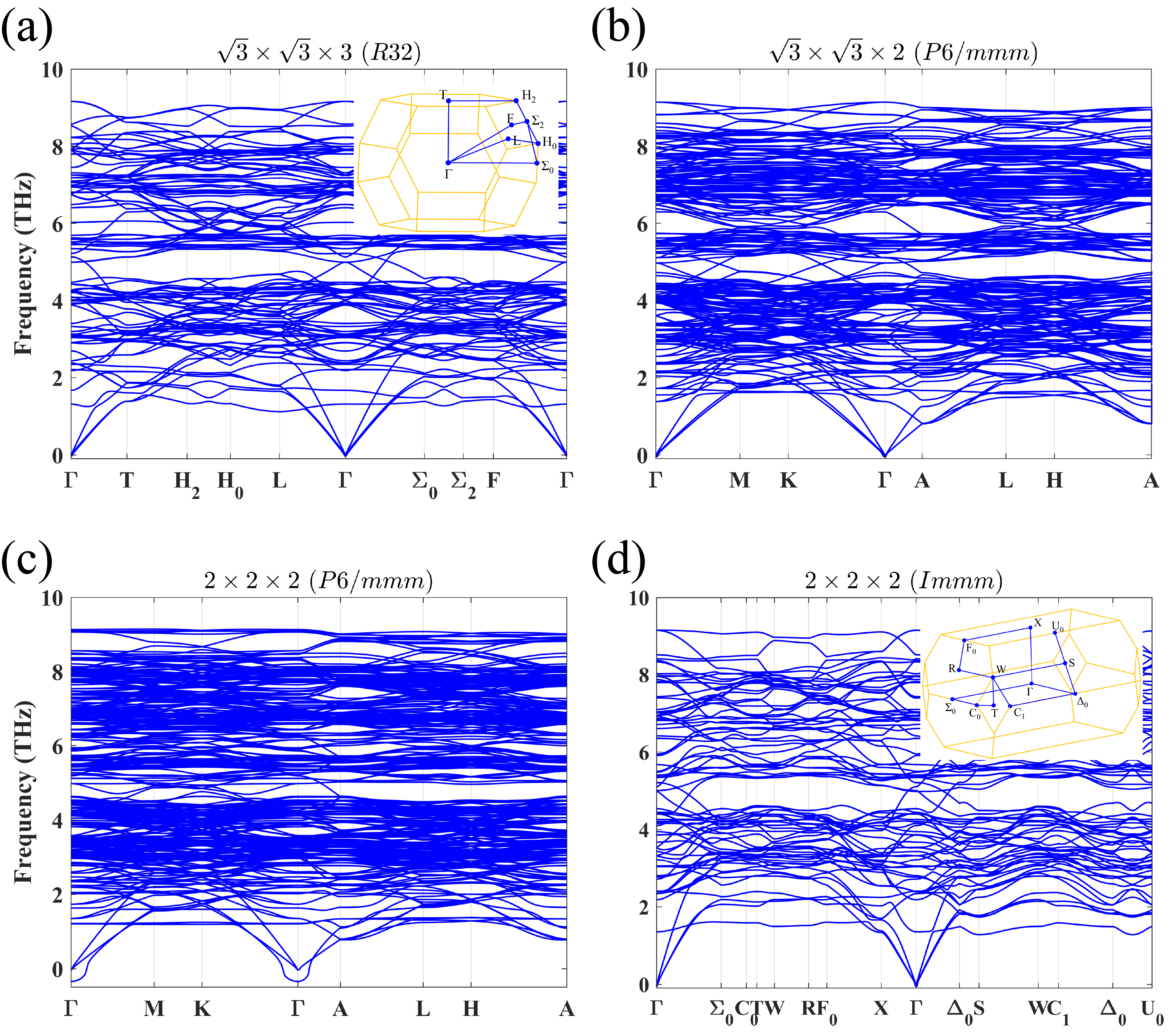}
\caption{ \label{FA1-PH_ALL} Full phonon band structures of all CDW structures of ScV$_6$Sn$_6$. The insets in (a) and (d) show the high-symmetry $k$-paths in the respective Brillouin zone, while the conventional $k$-paths are employed for the $P6/mmm$ structures in (b) and (c).
}
\end{figure}

\begin{figure}[tbp]
\includegraphics[width=0.95\linewidth]{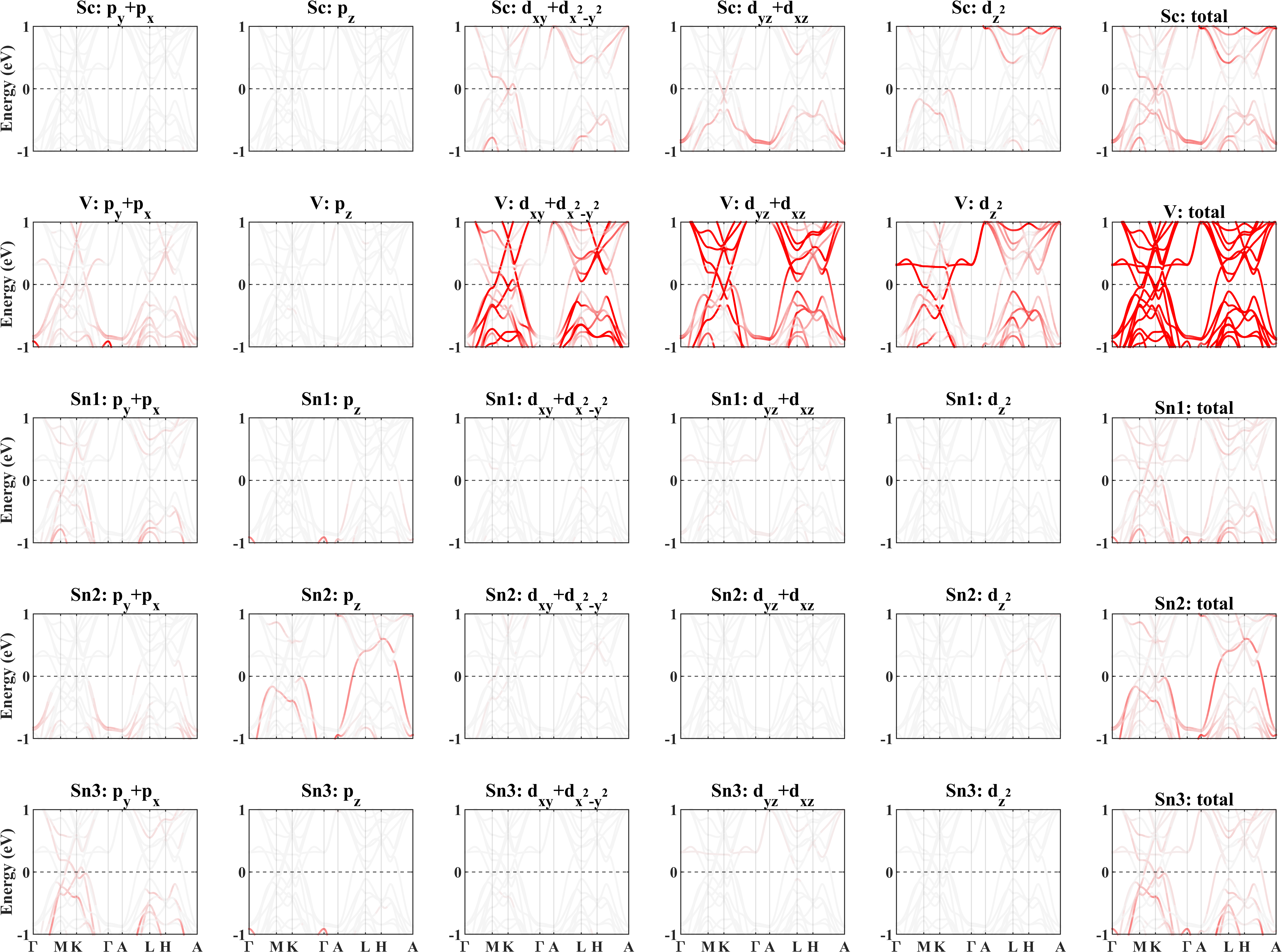}
\caption{ \label{FA1-PB_ALL} Atomic orbital resolved band structure of the pristine ScV$_6$Sn$_6$. The atom species are seen in Fig. 1 in the main text.
}
\end{figure}

\begin{figure}[tbp]
\includegraphics[width=0.7\linewidth]{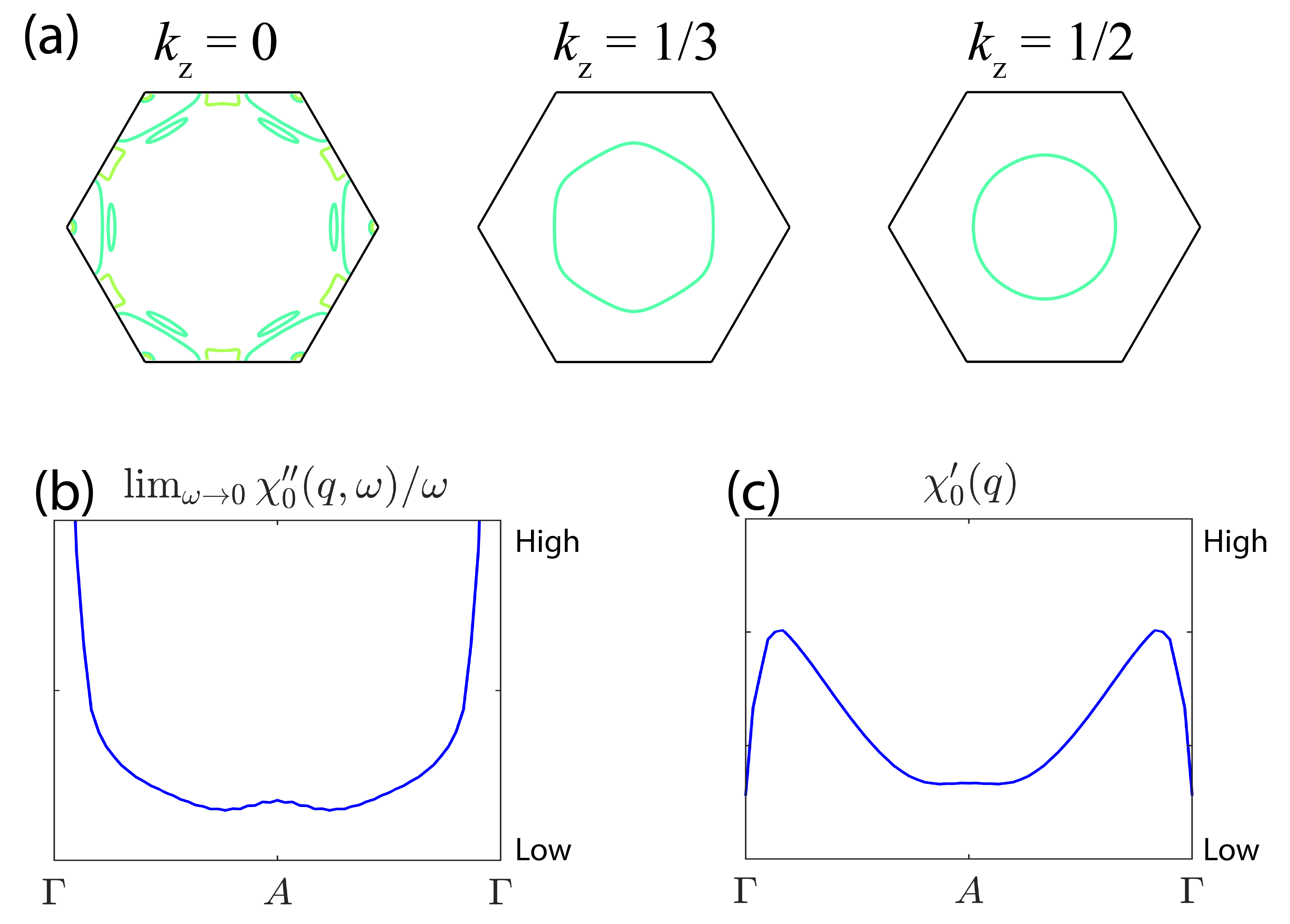}
\caption{ \label{FA1-FS} (a) Fermi surface slices on different $k_z$ planes. Fermi surfaces in the same color come from the same band. (b) and (c) are the imaginary and real part of the bare charge susceptibility of the pristine ScV$_6$Sn$_6$ along the out-of-plane direction.
}
\end{figure}

\begin{figure}[tbp]
\includegraphics[width=0.75\linewidth]{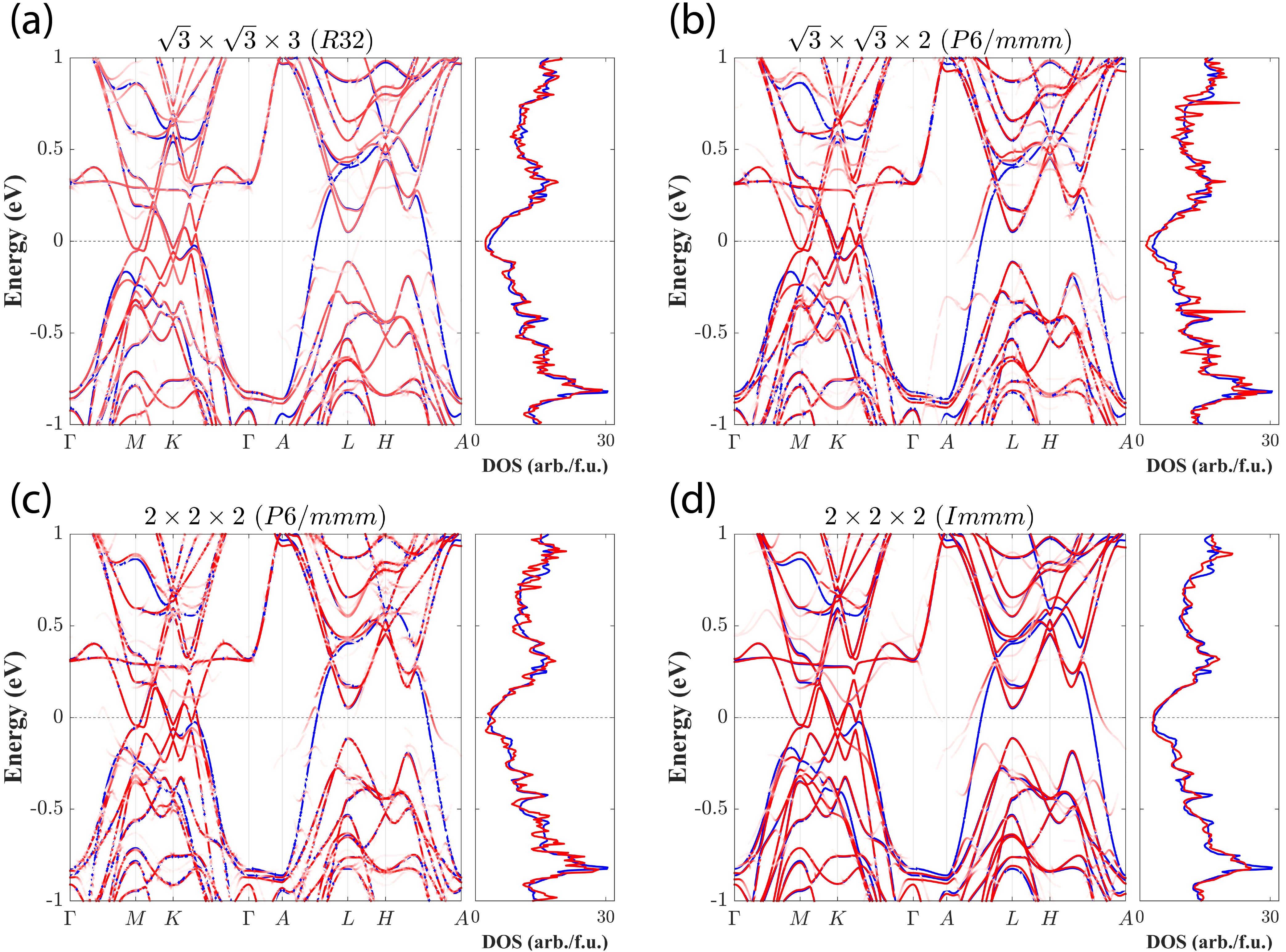}
\caption{ \label{Fig4-unfold} Unfolded band structures of CDW structures of ScV$_6$Sn$_6$, (a)-(d) for $\sqrt 3\times \sqrt 3 \times 3$ ($R32$), $\sqrt 3\times \sqrt 3 \times 2$ ($P6/mmm$), $2\times 2 \times 2$ ($P6/mmm$), and $2\times 2 \times 2$ ($Immm$), respectively.
The left part of each panel shows the band structure, where the weighted color red stands for the unfolded band structure of the CDW structure, and the blues curves represent the pristine phase band structure.
The right part of each panel shows the density of states (DOS), where the red and blue curves are for the CDW and pristine phases, respectively.
The blue curves in all panels are overlaid for convenience of comparison.
}
\end{figure}

\begin{figure}[tbp]
\includegraphics[width=0.95\linewidth]{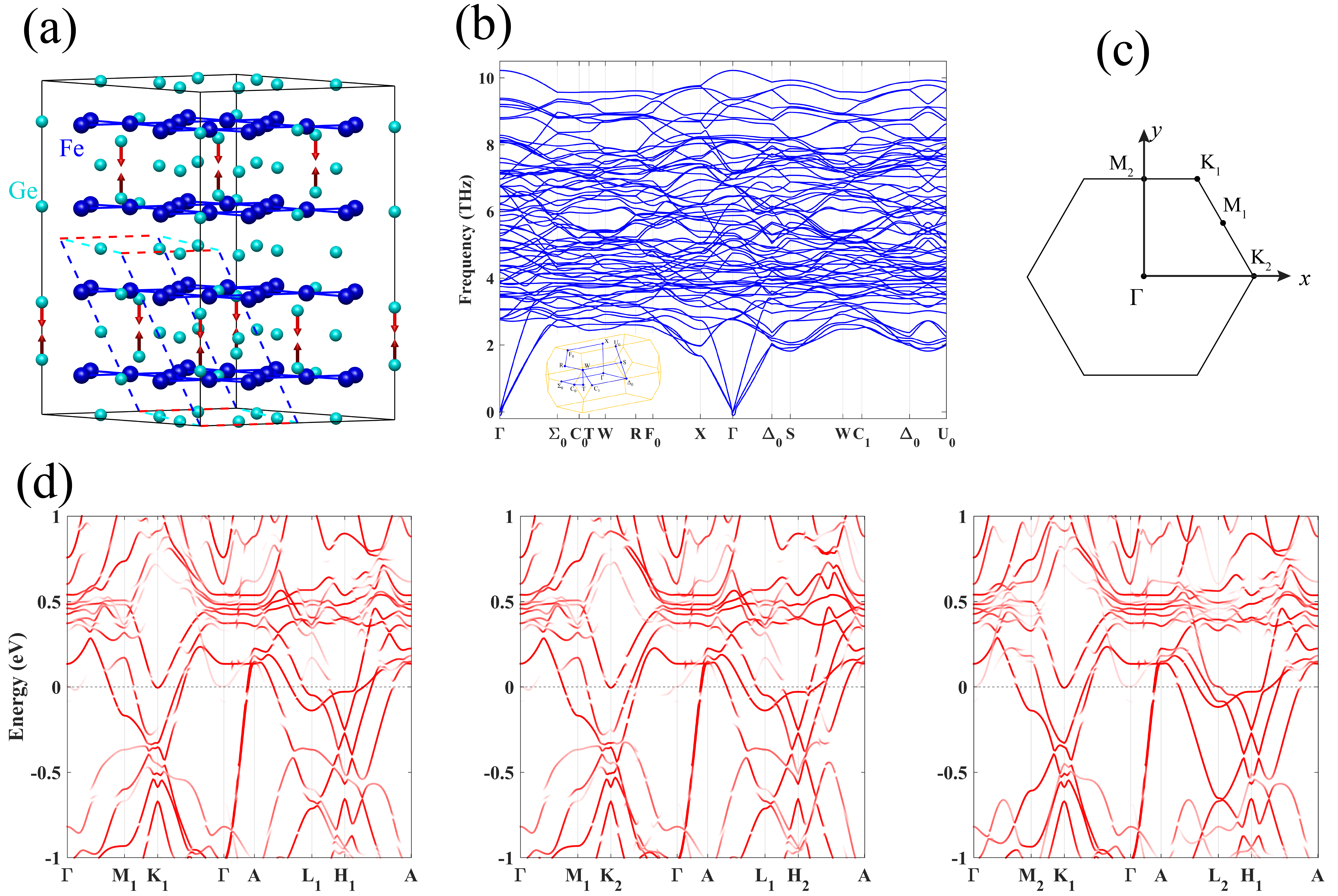}
\caption{Crystal and electronic structures of FeGe. (a) The CDW crystal structure of FeGe obtained from the $2 \times 2 \times 2$ ($Immm$) CDW structure of ScV$_6$Sn$_6$. The colored dashed rhombohedron indicates a primitive cell inside the hexagonal supercell of FeGe. The red vectors on atoms show the distortion pattern concerning the high-symmetry structure. (b) The phonon band structure of the structure in (a) shows the dynamic stability of the structure. The inset is the Brillouin zone showing the $ k$ path used in calculations. (c) is the in-plane Brillouin zone of the pristine FeGe. Due to symmetry breaking from $D_{6h}$ to $D_{2h}$, the inequivalent high-symmetry $k$ points are labeled. (d) The unfolded band structures of FeGe in the $Immm$ phase. The high-symmetry $k$ points are shown in (c) where points on the $k_z=0.5$ planes correspond to those in the $\Gamma-M-K$ plane.
}
\label{FA1}
\end{figure}

\newpage

%